\documentclass[prl,twocolumn,showpacs,amsfonts,amssymb,amsmath]{revtex4}
\usepackage{times}
\usepackage{txfonts}
\usepackage{graphicx}
\usepackage{psfrag}
\newsavebox{\LSIM}
\sbox{\LSIM}{{$\ \stackrel{\textstyle<}{\sim}\ $}}

\psfrag{tlabel}{$t$ ($\hbar/J$)}
\psfrag{yMIlabel}{$(\Delta n)_{MI}^2/n(n+1)$}
\psfrag{ylonglabel}{$[(\Delta n)^2(t,\epsilon)-(\Delta n)_{MI}^2(t)]/n(n+1)$}
\psfrag{eylabel}{$[U_{int}(\Delta n)-U_{int}(\Delta n=0)]/U/2$}
\psfrag{exlabel}{$(\Delta n)^2/n(n+1)$}
\psfrag{eylabelt}{$[U_{int}(\Delta n(\epsilon,t))-U_{int}(\Delta n(0,t)]/U/2$}
\begin{document}

\title{Probing States in the Mott Insulator Regime}
\author{D.C.~Roberts and K.~Burnett}
\affiliation{Clarendon Laboratory, Department of Physics, University of Oxford,
Oxford OX1~3PU, United Kingdom}
\date{\today}
\begin{abstract}
We propose a method to probe states in the Mott insulator regime produced from a condensate in an optical lattice.  We consider a system in which we create time-dependent number fluctuations in a given site by turning off the atomic interactions and lowering the potential barriers on a nearly pure Mott state to allow the atoms to tunnel between sites.  We calculate the expected interference pattern and number fluctuations from such a system and show that one can potentially observe a deviation from a pure Mott state.  We also discuss a method in which to detect these number fluctuations using time-of-flight imaging.
 \end{abstract}
 
\maketitle

Recently a quantum phase transition was induced in a condensate confined in an optical lattice\cite{munich}.  The condensate was observed to move from a state with long range phase coherence to an insulating state as  the depth of the periodic potential imposed on the atoms was slowly varied.  The disappearance of interference fringes implied the disappearance of long range phase coherence.  Furthermore, the fact that it was possible to regain coherence (in addition to the observation of a gap in the excitation spectrum) demonstrated that the Mott insulator phase had been achieved.  However, these observations do not give information on the precise nature of the state in the insulating regime.  Knowing the nature of the state is potentially important for  such applications as quantum computing \cite{qc} and Heisenberg-limited atom interferometry \cite{heisenberg}, both of which rely on having a precise number of atoms in each well.  In this paper, we propose a method to investigate the nature of the states in the insulating regime produced from condensates confined within an optical lattice. 

A condensate confined by an optical lattice, a series of interfering laser beams that creates a periodic potential, can be described by the Bose-Hubbard model \cite{jaksch}
\begin{equation}
\label{bosehubbard}
\hat H = -J \sum_{\langle i,j \rangle} \hat c^\dag_i \hat c_j + \frac{1}{2} U \sum_{i} \hat n_i(\hat n_i -1)-\mu \sum_{i} \hat n_i,
\end{equation} 
where the sum is taken over nearest neighbor sites.  Here, $J$ represents the coupling between nearest neighbor sites, $U$ is the interaction strength, $\mu$ is the chemical potential, and $\hat c^\dag_i$, $\hat c_i$, and $\hat n_i (= \hat c^\dag_i  \hat c_i$) are the atomic creation, annihilation and number operators for bosons at a lattice site $i$.  This model should be a sufficiently accurate description if the range of interactions is much smaller than the lattice spacing and the atoms are loaded into the lowest vibrational state of each well.  Both conditions can be well satisfied in practice.  One can adjust the coupling term, $J$, experimentally by varying the intensity of the lasers in the optical lattice and one can adjust the interaction term, $U$, by using Feshbach resonances \cite{innouye,donley}. 

If the optical lattice is turned on adiabatically to an intensity which is high enough to eliminate tunneling (i.e. $J=0$), the result should be a product of number states which have equal occupation numbers, i.e.
\begin{equation}
|\psi_{MI} \rangle= \prod_{i=1}^{N_s} (n!)^{-1/2}(\hat c^\dag_i)^n|0\rangle. 
\end{equation}
For convenience, we refer to  $|\psi_{MI} \rangle$ as a ``pure'' Mott insulator state.
Here, the filling factor $n$ is an integer \cite{n} defined to be the number of particles, $N$, divided by the number of lattice sites, $N_s$.
$|\psi_{MI} \rangle$ is a product of localized perfectly number squeezed states, i.e. $(\Delta n_i)^2 \equiv \langle \hat n_i^2\rangle - \langle \hat n_i \rangle^2=0$.  Thus the relative phase between each site is undetermined, and $|\psi_{MI} \rangle$ cannot be described by a macroscopic wavefunction.

We do, however, expect that in experiments the resultant state may deviate from $|\psi_{MI} \rangle$.  Tunneling between lattice sites cannot fully be eliminated in realistic experiments.  This would lead to corrections to the insulating ground state of the form  
$|\psi_{pert}\rangle =|\psi_{MI} \rangle+\frac{J}{2 U}\sum_{\langle i,j \rangle} \hat c_i^\dag \hat c_j|\psi_{MI} \rangle+O\left(\left(\frac{J}{U}\right)^2\right)$.   Other effects such as flow, loss, and the violation of adiabaticity could lead to a final state with an admixture of excited states. 
Therefore, we may better represent the Mott state produced experimentally as having an unspecified deviation from $|\psi_{MI} \rangle$ such that 
\begin{equation}
|\psi(\epsilon) \rangle=A[|\psi_{MI}\rangle+\frac{1}{\sqrt{n(n+1)}}\sum_{\langle i,j \rangle} \epsilon_{i,j} \hat c_i^\dag \hat c_j|\psi_{MI} \rangle ].
\end{equation}
For simplicity, in this paper we assume the complex parameters that measure the deviation  from $|\psi_{MI} \rangle$ are the same, i.e. $\epsilon_{i,j}=\epsilon$, giving the following normalization: $A=(1+|\epsilon|^2 2 N_s)^{-1/2}$.   We also assume periodic boundary conditions and restrict ourselves to 1-d as our analysis remains similar  at higher dimensions. 

After creating $|\psi(\epsilon) \rangle$, we apply a phase variation $\Delta \theta$ across the lattice to obtain
\begin{eqnarray}
&&|\psi_{\theta}(\epsilon) \rangle=A[|\psi_{MI}\rangle+\frac{\epsilon}{\sqrt{n(n+1)}} \nonumber \\
&&\sum_{ i}( e^{i \Delta \theta t} \hat c_i^\dag \hat c_{i+1}|\psi_{MI} \rangle+ e^{-i \Delta \theta t} \hat c_i^\dag \hat c_{i-1}) |\psi_{MI} \rangle].
\end{eqnarray}
Since $\Delta \theta$ only affects the excited states, it provides a convenient experimentally adjustable parameter to detect $\epsilon$.  This becomes apparent in the time-dependence of the interference pattern discussed below.  

In this letter, we propose to find an experimental signature of a nonzero $\epsilon$ in order to probe the nature of the insulating state.   We start with $|\psi_{\theta}(\epsilon)\rangle$ as the initial state (produced by slowly turning up the optical lattice and applying a phase variation across the lattice as described above).  We then make the assumption that the atomic interactions are switched off ($U=0$ at $t=0$) since interactions tend to localize atomic wave functions, limiting the number fluctuations per site. We also assume that the lattice barriers are dropped to a specified finite value ($J$ is jumped up to $J_0$ at $t=0$) \cite{timescale}.   The atoms would then be able to tunnel between lattice sites producing time-dependence of the variance of the atom number, or atomic number squeezing ($\Delta n$) \cite{Orzel}, in each site.  This time-dependence will be used to measure $\epsilon$.  

Our method stands in direct contrast to the recent experiment \cite{collapse} which measured the collapse and revival of the macroscopic matter wave field.  Where in \cite{collapse} the tunneling is suddenly switched off to allow a state in the superfluid regime to evolve purely through atomic interactions, in our system it is the atomic interactions that are switched off, allowing tunneling to occur in a nearly pure Mott insulator state.
 
The dynamics of our system can be calculated in the Heisenberg picture using a generalized phonon approach to seek solutions of the form  
\begin{equation}
 \hat{c}_i(t)=\frac{1}{N_s} \sum_{{\bf k}} \sum_j e^{i[{\bf k} \cdot ({\bf r}_i -{\bf r}_j)-\omega_{{\bf k}}t]} \hat c_j(0),
\end{equation}
where ${\bf r}_i$ is the coordinate of site $i$ and the wave vector ${\bf k}$ in momentum space runs over the first Brillouin zone.  
This can be substituted into the Heisenberg equations of motion, 
\begin{equation}
i \hbar \dot{\hat{c}}_i(t)=-\mu \hat c_i(t)-J_0(\hat c_{i+1}+\hat c_{i-1}),
\end{equation}
 to obtain $\mu=-2J_0$ and the dispersion relation, 
\begin{equation}
\hbar \omega_{k}=2 J_0(1-\cos(k a)),
\end{equation}
where $a$ is the lattice spacing.  
      
First, we calculate the time-dependence of the matter wave intensity $\langle \hat n_O(t) \rangle$, which is associated with the interference pattern, arising from the tunneling atoms,  where  $\langle \hat n_O(t) \rangle \equiv \langle \psi_{\theta}(\epsilon)| \hat O^\dag (t) \hat O(t) |\psi_{\theta}(\epsilon)\rangle$, $\hat O(t)= {N_s}^{-1/2} \sum_{l}\hat c_l(t) e^{i \Delta \phi l}$,
and $\Delta \phi$ is the phase difference between lattice sites \cite{yoo}. Since  $\sum_i e^{i({\bf k}-{\bf k'}) \cdot {\bf r_i}}= N_s \delta_{{\bf k},{\bf k'}}$ and $\sum_{\bf k}e^{i{\bf k} \cdot ({\bf r_i}-{\bf r_j})}= N_s \delta_{i,j}$,  we obtain 
\begin{equation}
\label{interfere}
\langle \hat n_O(t)\rangle=n+4 A^2 \Re(\epsilon) \sqrt{n(n+1)} \cos (\Delta \phi-\Delta \theta t),
\end{equation}
where the only interference arises from the initial superposition in $|\psi_{\theta}(\epsilon)\rangle$ and the only time-dependence arises from the applied phase variation on the initial state.  Recent numerical simulations have shown that one can expect interference fringes to remain deep into the Mott insulating phase \cite{roth,Kashurnikov2002a}.  Measuring this interference pattern to detect $\epsilon$ would be difficult if $\epsilon$ is small.

If we assume there is no applied phase gradient on the initial state, the interference pattern in eq. (\ref{interfere}) is {\it time-independent} even though atoms are tunneling and creating time-dependent number fluctuations.    Similarly, in a Mach-Zehnder interferometer using correlated number states as inputs, phase shifts can only be detected in the number fluctuations \cite{holland}.   Therefore, we look for dynamic experimental signatures of $\epsilon$ by investigating the time-dependence of the number fluctuations, or number squeezing, for a given state.    

From the results above we calculate the time-dependence of the number squeezing for a given site $i$, $(\Delta n)^2$, using $|\psi_{\theta}(\epsilon) \rangle$ as an initial state, i.e. $(\Delta n)^2(t,\epsilon)=\langle \psi_{\theta}(\epsilon) | \hat n^2(t) | \psi (\epsilon)\rangle - (\langle \psi_{\theta}(\epsilon) | \hat n(t) | \psi (\epsilon)\rangle)^2$, to obtain
\begin{equation}
\label{nsp}
\begin{split}
({\Delta n})^2(t,\epsilon)=& \frac{n(n+1)}{N_s^3}\sum_{1,2,3,4}\delta_{2+4,1+3} \{1+  \\ &
\cos\Omega_{1,2,3,4} t[\alpha(\epsilon,n)+\beta_R(\epsilon,n)\cos (k_2 a-\Delta \theta t)+ \\ &
\gamma(\epsilon,n)(\cos(k_4- k_1)a +\cos((k_2- k_1)a-2 \Delta \theta t)]   \\ &
 +\beta_I(\epsilon,n,N_s)\cos(k_2 a -\Delta \theta t) \sin\Omega_{1,2,3,4} t \} 
\end{split}
\end{equation}
where we have used the shorthand notation $\Omega_{1,2,3,4}={\omega_{k_1}-\omega_{k_2}+\omega_{k_3}-\omega_{k_4}}$ and $\delta_{2+4,1+3}=\delta_{{k_2}+{k_4},{k_1}+{k_3}}$.  The coefficient terms are given by
\begin{equation}
\label{coefffirst}
\alpha(\epsilon,n)=\frac{4 A^2 |\epsilon|^2}{n(n+1)}-1,
\end{equation}
\begin{equation}
\label{betaR}
\beta_R(\epsilon,n)=\frac{-8 A^2 \epsilon_R (2n+1)}{\sqrt{n(n+1)}},
\end{equation}
\begin{equation}
\label{betaI}
\beta_I(\epsilon,n)=\frac{-8 A^2 \epsilon_I }{\sqrt{n(n+1)}},
\end{equation}
\begin{equation}
\label{coefflast}
\gamma(\epsilon)=-2 A^2 |\epsilon|^2,
\end{equation}
where $\epsilon$ has been divided up into its real and imaginary parts, i.e. $\epsilon=\epsilon_R+i \epsilon_I$.  Note also that the normalization term $A$ depends on $\epsilon$ and $N_s$.  If we set $\epsilon=0$ in eq. (\ref{nsp}) we obtain the time-dependent squeezing using $|\psi_{MI}\rangle$ as an initial state, i.e.
\begin{eqnarray}
&&(\Delta n)^2_{MI}(t) \equiv (\Delta n)^2(t,0) = \nonumber \\
&&\frac{n(n+1)}{N_s^3}\sum_{1,2,3,4}\delta_{2+4,1+3}(1-\cos\Omega_{1,2,3,4} t).
\end{eqnarray} 
\begin{figure}[ht]
\begin{center}
\includegraphics[width=\columnwidth]{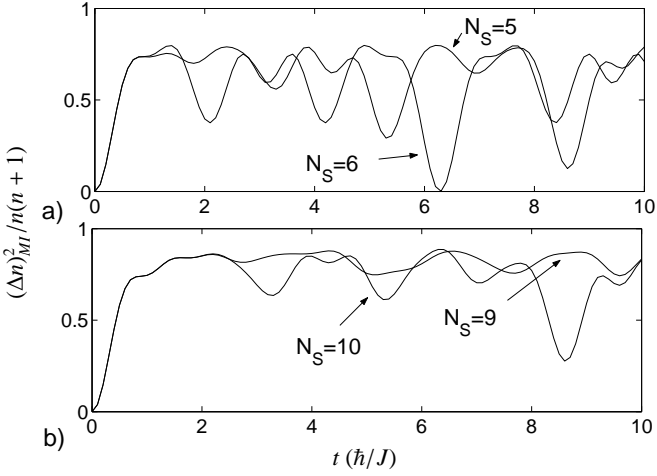}
\caption{\label{mifigure} The number fluctuations as a function of time in 1-d using the pure Mott state as an initial state plotted for different numbers of sites.}
\end{center}
\end{figure}
Figure \ref{mifigure} shows $(\Delta n)^2_{MI}(t)$ as a function of time for different number of lattice sites, $N_s$. The typical raising time in our system is approximately the tunneling time between two adjacent sites, i.e. $~\hbar/J$.

At $t=0$, $(\Delta n)^2(0,\epsilon)=4 A^2 |\epsilon^2|$ and $(\Delta n)^2_{MI}(0)=0$. 
As shown in eq. (\ref{interfere}), an interference experiment will only be sensitive to the difference between $(\Delta n)^2_{MI}$ and $(\Delta n)^2(t,\epsilon)$ at $t=0$ (if we ignore $\Delta \theta$).   Better results might be obtained by looking for different time varying signatures between $(\Delta n)^2(t,\epsilon)$ and $(\Delta n)^2_{MI}(t)$.   For simplicity, we set $\Delta \theta=0$ in the following analysis.

To detect an admixture to the pure Mott state characterized by $|\epsilon|<1$, we shall take advantage of the extrema  in $\beta_R(\epsilon,n)$ and $\beta_I(\epsilon,n)$.  

\begin{figure}[ht]
\begin{center}
\includegraphics[width=\columnwidth]{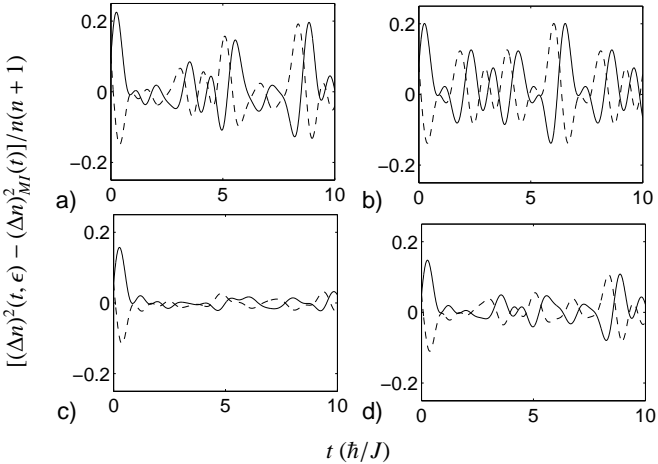}
\caption{\label{short} The time-dependence of the difference between the number  fluctuations starting with $|\psi(\epsilon) \rangle$ as an initial state and the number fluctuations starting with $|\psi_{MI} \rangle$ as an initial state.  We assume 1-d  with $\Delta \theta=0$, $n=1$ and a) $N_s=5$  b) $N_s=6$   c) $N_s=9$ and d) $N_s=10$.  The solid curve represents an initial state with $(\epsilon_R,\epsilon_I)=(0,(2 N_s)^{-1/2})$ and the dashed curve corresponds to  $(\epsilon_R,\epsilon_I)=(0,-(2 N_s)^{-1/2})$.}
\end{center}
\end{figure}

To detect a small $\epsilon_I$, we look to the behavior of $\beta_I(\epsilon,n,N_s)$ given in eq. (\ref{betaI}).  If $\beta_I(\epsilon,n,N_s)$ is at an extremum, i.e. $(\epsilon_R,\epsilon_I)=(0,\pm(2 N_s)^{-1/2})$, there should be a significant, and potentially observable, deviation from $(\Delta n)^2_{MI}(t)$ as shown in figure \ref{short}.  We can also see that for short times, this deviation only weakly depends on increasing $N_s$.  As you increase $n$, however, this deviation disappears as $[n(n+1)]^{-1/2}$ as seen in eq.~(\ref{betaI}).  Thus, increasing $n$ allows one to calibrate to the pure Mott insulator state ($note that (\Delta n)^2_{MI}(t) \propto (n(n+1)$).  

\begin{figure}[ht]
\begin{center}
\includegraphics[width=\columnwidth]{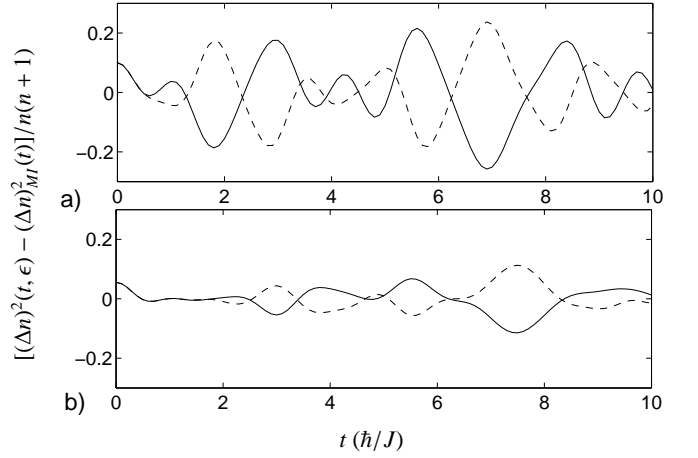}
\caption{\label{long} The time-dependence of the difference between the number  fluctuations starting with $|\psi(\epsilon) \rangle$ as an initial state and the number fluctuations starting with $|\psi_{MI} \rangle$ as an initial state.  We assume 1-d  with $\Delta \theta=0$, $n=1$ and a) $N_s=5$  and b) $N_s=9$.  The solid curve represents an initial state with $(\epsilon_R,\epsilon_I)=((2 N_s)^{-1/2},0)$ and the dashed curve corresponds to  $(\epsilon_R,\epsilon_I)=(-(2 N_s)^{-1/2},0)$.}
\end{center}
\end{figure}

To detect a small $\epsilon_R$, we look to the behavior of $\beta_R(\epsilon,n,N_s)$ given in eq. (\ref{betaR}). Figure \ref{long} shows that a potentially observable difference in the long time behavior of  $(\Delta n)^2(t,\epsilon)$ occurs at the extrema of $\beta_R(\epsilon,n)$, i.e. $(\epsilon_R,\epsilon_I)=(\pm(2 N_s)^{-1/2},0)$.  This effect is virtually independent of $n$ and is absent in our calculation if $N_s$ is even.  Also,  this deviation decreases with increasing $N_s$ so ideally the experiment to detect this deviation should be performed with a small number of lattice sites.  

Figures \ref{short} and \ref{long} only show the maximum deviation if $\epsilon$ is at one of the specified extrema.  As $\epsilon$ moves away from the extremum,  $(\Delta n)^2(t,\epsilon)$ gets closer to $(\Delta n)^2_{MI}(t)$.  However, the magnitude of the effects of small $\epsilon_I$ and $\epsilon_R$  depends on $N_s$.  Therefore, one can use different $N_s$ to amplify different $\epsilon$ to be measured. 

\begin{figure}[ht]
\begin{center}
\includegraphics[width=\columnwidth]{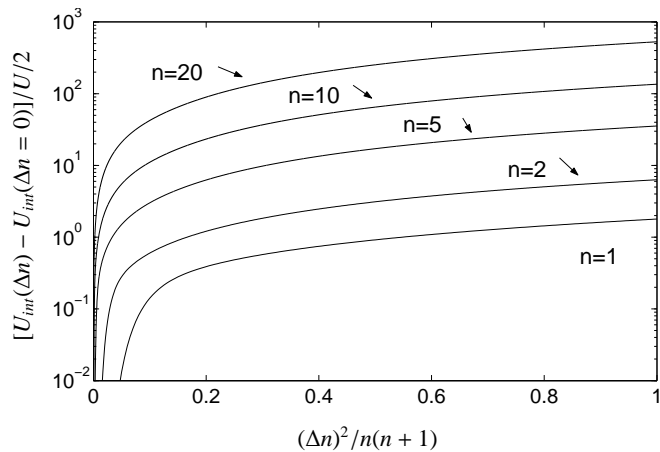}
\caption{\label{eint} The interaction energy converted from number fluctuations by rapidly turning off the tunneling and increasing interactions as a function of the number fluctuations for different filling factors.}
\end{center}
\end{figure}

These number fluctuations, or number squeezing ($\Delta n$), described above can potentially be measured using time-of-flight images.  To do this one rapidly turns off the tunneling, $J \rightarrow 0$, and increases the interactions $U$ (with the same timescales as described in \cite{timescale}), thus freezing $\Delta n$.  Each site would have a resulting interaction energy, $U_{int}$, that depends upon the number fluctuations before the tunneling was turned off and interactions increased.  This new interaction energy can be calculated from
\begin{equation}
U_{int} =\sum_{m=0}^{\infty} P(\Delta n, n, m) \frac{U}{2} m(m-1),
\end{equation}
where $P(\Delta n, n, m)$ is the probability that there are m particles in a site with $n$ mean number of atoms and a variance of $\Delta n$.
For a pure Mott insulator state where $\Delta n=0$ on each site, the interaction energy is given by
\begin{equation}
U_{int}(\Delta n=0)=\frac{U}{2} n(n-1).
\end{equation}
Assuming $P(\Delta n, n, m)$ is a Gaussian distribution we calculate $U_{int}(\Delta n)$ for various mean numbers as shown in Figure \ref{eint}.  

\begin{figure}[ht]
\begin{center}
\includegraphics[width=\columnwidth]{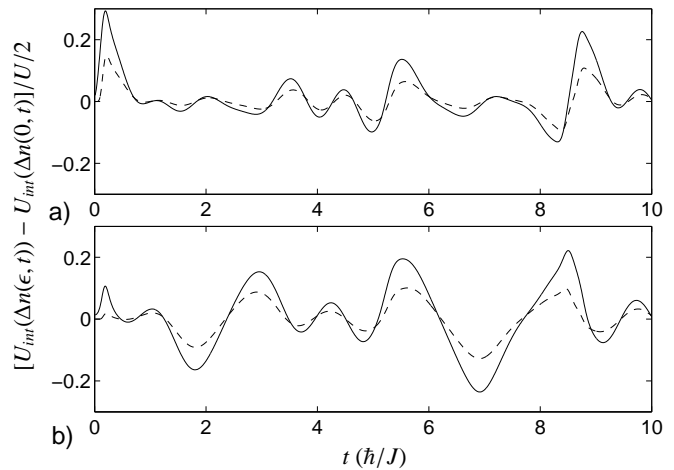}
\caption{\label{eint_t} The difference in the time-dependent interaction energy (resulting from converting from the time-dependent number fluctuations by rapidly turning off the tunneling and increasing interactions) as a result of finite $\epsilon$ assuming 1-d, $n=1$, and $N_s=5$.  In a) the solid curve represents an initial state with the extrema of $\beta_I(\epsilon,n,N_s)$, namely $(\epsilon_R,\epsilon_I)=((0,2 N_s)^{-1/2})=(0,0.3162)$, and the dashed curve corresponds to  $(\epsilon_R,\epsilon_I)=(0,0.1)$.  In b) the solid curve represents an initial state with the extrema of $\beta_R(\epsilon,n,N_s)$, namely $(\epsilon_R,\epsilon_I)=((2 N_s)^{-1/2},0)=(0.3162,0)$, and the dashed curve corresponds to  $(\epsilon_R,\epsilon_I)=(0.1,0)$.}
\end{center}
\end{figure}

In order for time-of-flight images to be sensitive to the interaction energy, $U_{int}$ must dominate the zero point energy for a given site, which occurs when   
 \begin{equation}
\label{zp}
na>>l_{lattice},
\end{equation}
where $a$ is scattering length and $l_{lattice}$ is the lattice spacing.
If eq. (\ref{zp}) is satisfied (which can be achieved by using Feshbach resonances to magnify $a$, having large filling factors  $n$, or a combination of both), then when the combined trapping potential is turned off, the interaction potential of the system $N_s U_{int}$ will be converted to kinetic energy and will consequently be measurable using time-of-flight imaging \cite{mewes,hollandjin}.  Thus one can deduce $\Delta n$.  In figure \ref{eint_t} we show the effect of $\epsilon$ on the time-dependence of $U_{int}$.

An alternative way to measure $\Delta n$, would be to bring the system adiabatically back through the Mott transition instead of by releasing the atoms as discussed for the two-mode case in \cite{inter}. Other possibilities to measure number fluctuations is by an ensemble of interference experiments or using a detector that detects pairs of atoms (perhaps using photoassociation).

In summary, we have described a method to probe for the presence of a deviation from a pure Mott state by investigating the time-dependence of the number fluctuations in a given site.  We have also proposed to detect these number fluctuations using time-of-flight measurements.  The effects of the deviation, $\epsilon$, are most easily observed if the number of lattice sites is small.   Low numbers of lattice sites can be isolated in large experiments (such as \cite{munich}), so that the insulating states produced can be investigated.  The detection of $\epsilon$ could be further verified by using the same method presented in this letter with different phase differences $\Delta \theta$ in the initial state. We expect that one could use this method of measuring $\epsilon$ to investigate the time-dependent formation of the Mott insulator state as well as to gain a better idea of how close to creating a pure Mott Insulator state one is able to come.  

The authors wish to thank William D. Phillips, Thomas Gasenzer, Robert Roth, and Jacob Dunningham for valuable discussions and Mark Lee for help with the numerical work.  This work was financially supported by the British Marshall Scholarship (D.C.R.) and the United Kingdom Engineering and Physical Sciences Research Council.


\begin{thebibliography}{9}
\bibitem{n} Commensurate fillings can be achieved in Mott domains in the inner region of the trap \cite{monte}.
\bibitem{timescale} The timescale for switching off the interactions and jumping up the tunneling is fast compared with the tunneling time between adjacent wells, but slow compared with the spacing between energy levels such that atoms remain in the vibrational ground state.  Such timescales can be achieved in experiments \cite{munich}.
\end{thebibliography}
\end{document}